\let\ifarxiv=\iftrue     
\pdfoutput=1
{\def\usepackage{ws-procs9x6}}

\ifarxiv
\documentclass[12pt,a4paper]{article}

\usepackage[nosort]{cite}
\usepackage[bulletsep]{collref}

\usepackage{amsmath,amssymb}

\usepackage[a4paper,text={490pt,650pt},centering]{geometry}

\usepackage[font=small,labelfont=bf,width=0.85\textwidth]{caption}

\makeatletter
\let\@pacs\@empty
\let\@keywords\@empty
\let\@preprint\@empty
\let\@authors\@empty
\let\@email\@empty
\let\@affiliation\@empty
\providecommand{\pacs}[1]{\gdef\@pacs{#1}}
\providecommand{\keywords}[1]{\gdef\@keywords{#1}}
\providecommand{\preprint}[1]{\toks@\expandafter{\@preprint#1\par}\edef\@preprint{\the\toks@}}
\renewcommand{\author}[1]{\ifx\@authors\@empty\toks@\expandafter{#1}\else\toks@\expandafter{\@authors, #1}\fi\edef\@authors{\the\toks@}}
\providecommand{\email}[1]{\ifx\@email\@empty\toks@\expandafter{#1}\else\toks@\expandafter{\@email, #1}\fi\edef\@email{\the\toks@}}
\providecommand{\affiliation}[1]{\gdef\@affiliation{#1}}

\usepackage{verbatim}
\newwrite\abs@out
\let\theabstract\@empty
\renewenvironment{abstract}
{%
  \immediate\openout\abs@out\jobname.abs%
  \@bsphack\let\do\@makeother\dospecials%
  \catcode`\^^M\active%
  \def\verbatim@processline{\immediate\write\abs@out{\the\verbatim@line}}%
  \verbatim@start
}%
{%
  \immediate\closeout\abs@out%
  \gdef\theabstract{\input{\jobname.abs}}%
  \@esphack%
}

\makeatother

\fi

\ifarxiv\else

\PassOptionsToClass{aps,prl}{revtex4-1}
\iftrue
\PassOptionsToClass{showpacs,showkeys,preprintnumbers}{revtex4-1}
\fi
\documentclass[10pt,letterpaper,twocolumn,amsmath,amssymb,final]{revtex4-1}

\makeatletter
\let\o@a@f\@author@finish
\def\@author@finish{\o@a@f%
\let\@authors\empty\def\AF@opr##1{}\def\CO@opr##1##2##3{}%
\def\AU@opr##1##2##3{\ifx\@authors\@empty\toks@\expandafter{##2}\else\toks@\expandafter{\@authors, ##2}\fi\edef\@authors{\the\toks@}}
\@AAC@list}
\makeatother

\fi

\ifnum\pdfoutput=1\else
\PassOptionsToPackage{hypertex}{hyperref}
\PassOptionsToPackage{draft}{graphicx}
\usepackage{showkeys}
\fi

\usepackage[bookmarks=true,hyperfigures=true]{hyperref}
\usepackage{graphicx}

\let\oldbfseries=\bfseries
\let\oldmdseries=\mdseries
\let\oldnormalfont=\normalfont
\renewcommand{\bfseries}{\oldbfseries\boldmath}
\renewcommand{\mdseries}{\oldmdseries\unboldmath}
\renewcommand{\normalfont}{\oldnormalfont\unboldmath}

\allowdisplaybreaks[3]

\providecommand{\hypersetup}[1]{}
\providecommand{\texorpdfstring}[2]{#1}

\hypersetup{plainpages=false}
\hypersetup{pdfpagemode=UseNone}
\hypersetup{bookmarksnumbered=true}
\hypersetup{pdfstartview=FitH}
\hypersetup{colorlinks=false}
\hypersetup{citebordercolor={.5 1 .5}}
\hypersetup{urlbordercolor={.5 1 1}}
\hypersetup{linkbordercolor={1 .7 .7}}

\DeclareMathSymbol{\Gamma}{\mathalpha}{letters}{"00}
\DeclareMathSymbol{\Delta}{\mathalpha}{letters}{"01}
\DeclareMathSymbol{\Theta}{\mathalpha}{letters}{"02}
\DeclareMathSymbol{\Lambda}{\mathalpha}{letters}{"03}
\DeclareMathSymbol{\Xi}{\mathalpha}{letters}{"04}
\DeclareMathSymbol{\Pi}{\mathalpha}{letters}{"05}
\DeclareMathSymbol{\Sigma}{\mathalpha}{letters}{"06}
\DeclareMathSymbol{\Upsilon}{\mathalpha}{letters}{"07}
\DeclareMathSymbol{\Phi}{\mathalpha}{letters}{"08}
\DeclareMathSymbol{\Psi}{\mathalpha}{letters}{"09}
\DeclareMathSymbol{\Omega}{\mathalpha}{letters}{"0A}

\newcommand{\gen}[1]{\mathfrak{#1}}
\newcommand{\genY}[1]{\widehat{\mathfrak{#1}}}
\newcommand{\superN}{\mathcal{N}}

\newcommand{\alg}[1]{\mathfrak{#1}}
\newcommand{\grp}[1]{\mathrm{#1}}

\newcommand{\del}{\partial}

\newcommand{\Amp}{\mathcal{A}}

\newcommand{\Complex}{\mathbb{C}}

\ifx\genfrac\sdflkaj\else\fi

\newcommand{\lrbrk}[1]{\left(#1\right)}
\newcommand{\bigbrk}[1]{\bigl(#1\bigr)}

\newcommand{\comm}[2]{[#1,#2]}

\newcommand{\sprod}[2]{\langle#1,#2\rangle}

\newcommand{\nln}{\nonumber\\}
\newcommand{\eqtab}{&}
\newcommand{\eqcr}{\\}

\def\[{\begin{equation}}
\def\]{\end{equation}}
\def\<{\begin{eqnarray}}
\def\>{\end{eqnarray}}

\makeatletter
\def\mr@ignsp#1 {\ifx\:#1\@empty\else #1\expandafter\mr@ignsp\fi}%
\newcommand{\multiref}[1]{\begingroup
\xdef\mr@no@sparg{\expandafter\mr@ignsp#1 \: }%
\def\mr@comma{}%
\@for\mr@refs:=\mr@no@sparg\do{\mr@comma\def\mr@comma{,}\ref{\mr@refs}}%
\endgroup}
\makeatother

\newcommand{\hypref}[2]{\ifx\href\asklfhas #2\else\href{#1}{#2}\fi}

\renewcommand{\eqref}[1]{(\multiref{#1})}

\makeatletter
\newlength{\apb@width}
\newcommand{\autoparbox}[2][c]{\settowidth{\apb@width}{#2}\parbox[#1]{\apb@width}{#2}}

\makeatother

\ifx\href\asklfhas\newcommand{\href}[2]{#2}\fi
\newcommand{\arxivlink}[1]{\href{http://arxiv.org/abs/#1}{arxiv:#1}}

\begin{document}

\preprint{\arxivlink{1103.0646}}
\preprint{AEI-2011-006}

\title{\ifarxiv\texorpdfstring{$\mathfrak{B}$}{B}\else B\fi onus Yangian Symmetry for the 
\ifarxiv\texorpdfstring{\\}{}\fi
Planar S-Matrix of \texorpdfstring{$\superN=4$}{N=4} Super Yang-Mills}

\author{Niklas \ifarxiv\texorpdfstring{$\mathfrak{B}$}{B}\else B\fi eisert}
 \email{nbeisert@aei.mpg.de}

\author{\ifarxiv\texorpdfstring{$\mathfrak{B}$}{B}\else B\fi urkhard U.\ W.\ Schwab}
 \email{buws2@aei.mpg.de}

\affiliation{%
Max-Planck-Institut f\"ur Gravitationsphysik%
\ifarxiv\\\else\ (\fi%
Albert-Einstein-Institut%
\ifarxiv\\\else)\ \fi%
Am M\"uhlenberg 1, 14476 Potsdam, Germany%
}

\date{\today}

\begin{abstract}
Recent developments in the determination of the planar 
S-matrix of $\mathcal{N}=4$ Super Yang-Mills 
are closely related to its Yangian symmetry. 
Here we provide evidence for a yet unobserved additional symmetry:
the Yangian level-one helicity operator.
\end{abstract}


\ifarxiv

\makeatletter
\thispagestyle{empty}
\begin{flushright}
\footnotesize\ttfamily\@preprint
\end{flushright}
\vspace{0.5cm}

\begin{centering}
\begingroup\Large\bfseries\@title\par\endgroup
\vspace{1cm}

\begingroup\scshape\@authors\par\endgroup
\vspace{5mm}

\begingroup\itshape\@affiliation\par\endgroup
\vspace{3mm}

\begingroup\ttfamily\@email\par\endgroup
\vspace{1cm}

\textbf{Abstract}\vspace{7mm}

\begin{minipage}{12.7cm}
\theabstract
\end{minipage}

\end{centering}

\makeatother

\vspace{1cm}
\hrule height 0.75pt
\vspace{1cm}


\fi

\ifarxiv\else

\maketitle

\fi


\makeatletter
\hypersetup{pdftitle={\@title}}%
\hypersetup{pdfsubject={PACS numbers: \@pacs}}%
\hypersetup{pdfkeywords={\@keywords}}%
\hypersetup{pdfauthor={\@authors}}%
\makeatother

\section{Introduction}

The S-matrix is an object of central importance to
quantum field theories.
Unfortunately it is extremely challenging to compute it
because we typically have to rely on perturbative methods.
Symmetries are extremely helpful in constraining the result.
A couple of years ago it has been noticed that loop scattering amplitudes
in \emph{maximally supersymmetric} Yang-Mills ($\superN=4$ SYM) theory
in the \emph{planar limit} 
are simpler than one might expect \cite{Anastasiou:2003kj,Bern:2005iz}
based on the known symmetries such as superconformal symmetry.
This simplicity was traced to a hidden superconformal symmetry
of dual Feynman graphs \cite{Drummond:2006rz,Drummond:2008vq,Brandhuber:2008pf}.
These observations have sparked enormous progress in the determination 
of the planar S-matrix in this model,
see \cite{Alday:2008yw,Drummond:2010km} for reviews.

The appearance of \emph{dual symmetries} is well understood
from the point of view of the AdS/CFT dual string theory
where ordinary and dual symmetries are exchanged by a 
T-self-duality \cite{Alday:2007hr,Beisert:2008iq,Berkovits:2008ic}.
The closure of these two sets of symmetries forms 
an infinite-dimensional algebra \cite{Beisert:2009cs}
known from the context of \emph{planar integrability} \cite{Beisert:2010jr}.
Indeed, it was shown that the planar tree-level S-matrix
in $\superN=4$ SYM is invariant under the
Yangian $\grp{Y}(\alg{psu}(2,2|4))$ \cite{Drummond:2009fd}
for the superconformal algebra $\alg{psu}(2,2|4)$.

In this letter we find evidence for an additional symmetry 
of planar scattering amplitudes.
It is not part of the above Yangian, but it fits nicely into its structure:
It is the level-one recurrence $\genY{B}$ of the $\alg{u}(1)$ outer automorphism $\gen{B}$ of $\alg{psu}(2,2|4)$.
The automorphism, sometimes called ``bonus symmetry'' \cite{Intriligator:1998ig}, 
counts the helicity of particle states. We will refer to it 
as the \emph{hypercharge}.
It is clear that helicity is generally not conserved in scattering amplitudes%
\ifarxiv\ (MHV: ``Maximum Helicity \emph{Violating}'')\fi, 
but here we will argue that the level-one hypercharge $\genY{B}$
is indeed a proper symmetry.

An analogous observation has been made in the context of 
the worldsheet S-matrix \cite{Beisert:2005tm} for planar $\superN=4$ SYM.
This S-matrix is based on an extension of $\alg{psu}(2|2)$
which also possesses a $\alg{u}(1)$ automorphism.
It was shown to be exactly symmetric under the Yangian level-one 
automorphism \cite{Matsumoto:2007rh,Beisert:2007ty},
sometimes called ``secret symmetry'',
while there seem to be obstacles for the other levels.

In the following we shall present evidence in favor of 
the Yangian level-one hypercharge $\genY{B}$ 
being a symmetry of the planar S-matrix:
As a first check we show compatibility with the cyclic nature
of color-ordered amplitudes.
Next we confirm explicitly that tree-level MHV amplitudes are invariant.
We will also check that the Grassmannian formula for leading singularities \cite{ArkaniHamed:2009dn}
respects our symmetry, and hence symmetry extends to all tree-amplitudes at least.
Finally, we show that our symmetry becomes exact in a distributional sense
when appropriate correction terms are added. 

\section{Setup}

In the planar limit the S-matrix is described by color ordered 
scattering amplitudes $\Amp_n$.
In $\superN=4$ SYM the particle momentum, flavor and helicity 
are conveniently encoded by a spinor-helicity superspace:
The amplitude $\Amp_n$ 
is a function of the $\Lambda_k=(\lambda_k,\tilde\lambda_k,\eta_k)$, $k=1,\ldots,n$.
The complex conjugate spinors $\lambda_k,\tilde\lambda_k\in\Complex^2$ 
describe a real massless momentum $p_k=\lambda_k\tilde\lambda_k$.
Likewise, $\eta_k\in\Complex^{0|4}$ is a Grassmann variable to encode flavor.

Free superconformal symmetries $\gen{J}^A$ are represented on particles
by suitable differential operators $\gen{J}^A_k$ acting on $\Lambda_k$ \cite{Witten:2003nn}.
As usual they act on an amplitude $\Amp_n$ as a sum over all external particles. 
Conversely, the Yangian symmetries $\genY{J}^A$
act on pairs of external particles as follows
\[\label{eq:SCYangDef}
\gen{J}^A = \sum_{i=1}^n\gen{J}_i^A,
\qquad
\genY{J}^A = f^{A}_{BC}\sum_{j<k=1}^n\gen{J}_j^B\gen{J}_k^C.
\] 
Here $f$ denotes the $\alg{psu}(2,2|4)$ structure constants.
When promoting them to $\alg{u}(2,2|4)$ we find the free action of 
the Yangian level-one hypercharge $\genY{B}$
\[\label{eq:Bhdef}
\genY{B}
=
\sum_{k=1}^{n-1}\sum_{j=k+1}^n 
\lrbrk{\vphantom{\bar{\gen{S}}^{B,}_j }
\ifarxiv\else\begin{array}{l}\fi
\gen{Q}_k^{\alpha b} \gen{S}_{j,\alpha b} 
-\bar{\gen{Q}}_{k,b}^{\dot{\alpha}}\bar{\gen{S}}_{j,\dot{\alpha}}^b
\ifarxiv\else\eqcr[0.3em]\qquad\fi
-\gen{Q}_j^{\alpha b} \gen{S}_{k,\alpha b} 
+\bar{\gen{Q}}_{j,b}^{\dot{\alpha}}\bar{\gen{S}}_{k,\dot{\alpha}}^b
\ifarxiv\else\end{array}\fi
}.
\]
Here $\gen{Q}$ and $\gen{S}$ denote the superconformal 
translations and boosts, respectively.
Gladly, the broken hypercharge $\gen{B}= \sum_i \eta^A_{i} \partial/\partial\eta^A_{i}$ 
does not appear in $\genY{B}$.
In the following we will argue that 
the bonus Yangian generator $\genY{B}$ defined in 
\eqref{eq:Bhdef} is a symmetry of the planar S-matrix.

\section{Cyclicity}

Color-ordered amplitudes are invariant under cyclic shifts of the external particles.
Symmetries have to respect this property. 
The generators $\gen{J}^A$ in \eqref{eq:SCYangDef} are cyclic,
whereas the $\genY{J}^A$ typically map cyclic functions to non-cyclic ones.
This can be seen by shifting the summation range in \eqref{eq:SCYangDef} by one unit
\cite{Drummond:2009fd}
\[\label{eq:cyclfree}
\genY{J}^A_{(2,n+1)} - \genY{J}^A_{(1,n)} 
= f^{A}_{BC}\gen{J}_1^{B}\gen{J}^{C} + f^{A}_{BC}f^{BC}_D \gen{J}_1^D.
\]
Luckily, for scattering amplitudes the r.h.s.\ is zero:
The first term vanishes because $\gen{J}^A$ is a symmetry,
and the second one because the dual Coxeter number of
$\alg{psu}(2,2|4)$ is zero.
Hence the Yangian generators are cyclic.

For the proposed bonus Yangian symmetry $\genY{B}$ the situation is slightly different:
The first term vanishes as before due to superconformal symmetry.
In $\alg{u}(2,2|4)$ the combination 
$f^{A}_{BC}f^{BC}_D$ is proportional to $\delta^{A}_{\gen{B}}\delta_{D}^{\gen{C}}$,
and we obtain an additional $\gen{C}_1$.
This vanishes because the central charge of all individual
particles is zero, and $\genY{B}$ is indeed cyclic.

\section{Invariance of MHV amplitudes}

First act with $\genY{B}$ in \eqref{eq:Bhdef}
on MHV amplitudes
\cite{Parke:1986gb,Berends:1987me}
\[
\Amp_{n,2}=
\frac{\delta^4(P)\,\delta^8(Q)}{\prod_j\sprod{j}{j+1}}\,,
\qquad 
\begin{array}{l}
P=\sum_j \lambda_j\tilde\lambda_j,\\
Q=\sum_j \lambda_j\eta_j,
\end{array}
\]
with the spinor product $\sprod{j}{k}:=\varepsilon_{\alpha\beta}\lambda^\alpha_{j}\lambda^\beta_k$.
The fermionic derivatives in $\gen{S}$, $\bar{\gen{Q}}$
act only on $\delta^8(Q)$, and we obtain
\begin{align}
\ifarxiv\else\!\!\fi
\gen{Q}_{j}^{\alpha b} \gen{S}_{k,\alpha b} \Amp_{n,2}
&= \lambda_j^\alpha\eta_j^b\frac{\del}{\del\lambda^\alpha_{k}}\lambda_k^\beta 
 \frac{\del \delta^8(Q)}{\del Q^{\beta b}}\,\frac{\delta^4(P)}{\prod_i\sprod{i}{i+1}}\,,
\label{mhvcalc}
\\
 \gen{\bar Q}_{k,b}^{\dot\alpha} \gen{\bar S}_{j,\dot\alpha}^b \Amp_{n,2}
&=-\eta_j^b\frac{\del}{\del\tilde\lambda^{\dot\alpha}_{j}}\tilde\lambda_{k}^{\dot\alpha}\lambda_k^{\beta} 
 \frac{\del \delta^8(Q)}{\del Q^{\beta b}}\,\frac{\delta^4(P)}{\prod_i\sprod{i}{i+1}}\,.
\label{mhvcalc2}
\end{align}
The bosonic derivative in \eqref{mhvcalc} acts on the $\lambda^\beta_k$, 
both delta functions and the spinor brackets in the denominator.
The action on $\delta^4(P)$ produces 
\[
\frac{\lambda_j^\alpha\eta_j^b\lambda_k^\beta \tilde\lambda_k^{\dot\alpha} }
{\prod_i\sprod{i}{i+1}}\,
\frac{\del\delta^8(Q)}{\del Q^{b \beta}}\,
\frac{\del\delta^4(P)}{\del P^{\alpha\dot\alpha}}
\] 
which cancels precisely the contribution from \eqref{mhvcalc2}.
The contribution from the action on $\delta^{(8)}(Q)$ 
is proportional to 
$\lambda_j^\alpha\eta_j^a\lambda_k^\beta\eta_k^b
\del^2\delta^8(Q)/\del Q^{a\beta}\del Q^{b \alpha}$.
This expression is symmetric in $j$ and $k$
and vanishes due to the antisymmetry of \eqref{eq:Bhdef}.
Next consider the contribution originating from the derivative acting on the spinor brackets.
Combining the contributions from $\partial/\partial\lambda_k$ and $\partial/\partial\lambda_{k+1}$ (after a shift $k\to k+1$)
acting both on the same $\sprod{k}{k+1}$ we obtain 
\[
-\lambda_j^\alpha\eta_j^a 
 \frac{\del \delta^8(Q)}{\del Q^{\alpha a}}\,\frac{\delta^4(P)}{\prod_i\sprod{i}{i+1}}\,.
\] 
This term cancels identically with the derivative acting on $\lambda^\beta_k$ in \eqref{mhvcalc}.
The shift $k\to k+1$ leaves behind two boundary terms in the sum \eqref{eq:Bhdef}.
Here we complete the sum $\sum_j\lambda_j\eta_j=Q$ and
move it past the derivative acting on $\delta^8(Q)$.
A careful calculation shows that all remaining terms cancel.
Thus we find that $\genY{B}$ leaves MHV amplitudes invariant, 
$\genY{B} \Amp_{n,2} = 0$.
 
\section{Invariance of the Grassmannian Integral}

We complete the proof of invariance of tree amplitudes under $\genY{B}$ 
using the Grassmannian integral formula \cite{ArkaniHamed:2009dn} 
for leading singularities $\mathcal{L}_{n,k}$ in N${}^{k-2}$MHV amplitudes
$\Amp_{n,k}$ with $2<k\leq n-2$
\[
\mathcal{L}_{n,k}\simeq\int \frac{d^{k\times n}t\,\prod_{a=1}^k \delta_a}{\mathcal{M}_1(t)\cdots \mathcal{M}_n(t)}
\,,
\quad
\delta_a=\delta^{4|4}(\textstyle\sum_{j=1}^n t_{aj}\mathcal{Z}_j).
\]
Here $t$ is a $k\times n$ matrix, 
and $\mathcal{M}_a$ represent its minors of $k$ consecutive rows starting at $a$.
The particle momenta are encoded using supertwistors
$\mathcal{Z}^{\mathcal{A}} = (\tilde\mu^{\alpha},\tilde\lambda^{\dot\alpha},\eta^{a})$
with $\tilde \mu$ the Fourier conjugate to $\lambda$ \cite{Witten:2003nn}
(the calculation using momentum twistors \cite{Mason:2009qx} is virtually the same).
The $\alg{u}(2,2|4)$ algebra is now represented on particles by linear differential operators 
$\gen{J}^{\mathcal{A}}{}_{\mathcal{B}}=(-1)^{\mathcal{B}}\mathcal{Z}^{\mathcal{A}}\partial_{\mathcal{B}}$.
The corresponding Yangian generators take the form
\[\label{eq:Ybifund}
\genY{J}^{\mathcal{A}}{}_{\mathcal{B}}
=
\genY{J}^{\mathcal{A}}_{<,\mathcal{B}}-\genY{J}^{\mathcal{A}}_{>,\mathcal{B}},
\qquad
\genY{J}^{\mathcal{A}}_{\lessgtr,\mathcal{B}}
=
\sum_{j\lessgtr k=1}^n
\gen{J}^{\mathcal{A}}_{j,\mathcal{C}}
\gen{J}^{\mathcal{C}}_{k,\mathcal{B}}
.
\]
In this form our generator reads
$\genY{B}=\genY{J}^{\mathcal{A}}{}_{\mathcal{A}}$.

We shall now show that both contributions $\genY{B}_{<}\simeq \genY{B}_{>}\simeq 4k(k-1)$ 
on $\mathcal{L}_{n,k}$, and hence $\genY{B}$ annihilates the Grassmannian integral.
Our proof and notation follows along the lines of \cite{Drummond:2010qh}, 
where the calculation for all previously known Yangian operators can be found in detail. 
We apply $\genY{B}_<$ to $\mathcal{L}_{n,k}$ and obtain 
\[
\sum_{b=1}^k 
\int \frac{d^{k\times n}t}{\mathcal{M}_1\cdots\mathcal{M}_n}\,
(-1)^\mathcal{A}[\mathcal{O}_b^\mathcal{A}-\mathcal{V}_b^\mathcal{A}]
(\del_\mathcal{A}\delta_b)\prod_{a\neq b}\delta_a.
\]
We have defined 
$\mathcal{O}_b^{\mathcal{A}}:=\sum_{a,i<j}\mathcal{Z}^{\mathcal{A}}_i t_{ai} (\partial/\partial t_{aj})t_{bj}$
and $\mathcal{V}_b^{\mathcal{A}}:=\sum_{i<j}\mathcal{Z}^{\mathcal{A}}_i t_{bi}$.
Now we commute the operator $\mathcal{O}_b^{\mathcal{A}}$ 
past the minors $\mathcal{M}_p$ in the denominator. 
At this point it is important to be very careful as to not overlook 
the contributions that arise from the supertrace
over the index $\mathcal{A}$,
cf.\ footnote 9 in \cite{Drummond:2010qh}:
Specifically due to the wrapping of the minors $\mathcal{M}_p$ around 
the end of the $k\times n$ matrix $t_{bj}$ 
it is necessary to make a distinction between the two cases $p\leq n-k+1$ and $p>n-k+1$. 
In the first case the supertrace has no impact on the calculation. 
In the latter case, however, it is inevitable to use the constraint 
from the delta functions $\delta_a$ twice. 
For the delta function bearing the derivative $\del_\mathcal{A}\delta_b$
the supertrace leaves an extra term proportional 
to the Grassmannian integral.
The result of this operation is given by
\begin{align}
&(-1)^{\mathcal{A}}\sum_{b=1}^k \sum_{p=n-k+2}^n \sum_{s=p}^n \sum_{i=1}^{s-1} 
\frac{1}{\mathcal{M}_p} \mathcal{Z}_i^\mathcal{A} t_{bs}
\mathcal{M}_p^{s\to i}(\del_\mathcal{A}\delta_b)
\ifarxiv\else\nln\eqtab\quad\fi
 = - \sum_{s\geq p=n-k+2}^n \left[ 8 \delta_b + (-1)^\mathcal{A}
\sum_{b=1}^k \mathcal{Z}_s^\mathcal{A} t_{bs} (\del_\mathcal{A}\delta_b)\right]\label{addterm}
\end{align} 
after adding and subtracting the terms missing to make the sum over $s$ 
range from $1$ to $n$ and then performing the partial integration 
of the derivative $\del_\mathcal{A}$. 
The form of the second part on the right hand side follows 
from the antisymmetry of the minors which singles out the terms with $i=s$. 
The two sums in the first term evaluate straightforwardly to $-4k(k-1)$. 
Repeating this procedure for the second term on the right hand side 
of \eqref{addterm} yields a factor of $8k(k-1)$, 
such that in the end one is left with $4k(k-1) \mathcal{L}_{n,k}$. 
The contributions from $\genY{B}_<$ and $\genY{B}_>$ cancel 
each other leaving only a total derivative under the integral 
as in the proof of \cite{Drummond:2010qh}. 
This confirms that $\genY{B}$ is a symmetry of all leading singularities 
and, in particular, of tree level amplitudes of $\mathcal{N}=4$ SYM. 

\section{Distributional Contributions}

Due to the holomorphic anomaly
$(\partial/\partial\bar\lambda^{\dot\alpha})
\sprod{\lambda}{\mu}^{-1}
=2\pi\varepsilon_{\dot\alpha\dot\beta}\bar\mu^{\dot\beta}\delta^2(\sprod{\lambda}{\mu})$
the above derivations disregard certain distributional contributions
which at first sight violate the exactness of the symmetry. 
In \cite{Bargheer:2009qu} it was shown that 
in the case of superconformal boosts $\gen{S}$, $\bar{\gen{S}}$
the representation can be corrected to restore the symmetry.
The correction terms are operators 
$\gen{S}^+$, $\bar{\gen{S}}^+$ 
which act on an amplitude with $n-1$ legs
and return an amplitude with $n$ legs.
The statement of exact invariance then takes the form
$\gen{S}\Amp_n+\gen{S}^+\Amp_{n-1}=0$.
As our generator $\genY{B}$ contains the superconformal boosts
we will have to correct it by a suitable length-changing deformation
$\genY{B}^+$ such that
\[\label{eq:ExactSym}
\genY{B}\Amp_n+\genY{B}^+\Amp_{n-1}=0.
\]

Before we consider the correction, let us briefly discuss how 
to work with length-changing operators.
The correction $\gen{S}^+$ acts on an $(n-1)$-particle function
and generates an $n$-particle function.
We define the action on the first leg via a three-vertex $S^+$
\cite{Beisert:2010gn}
\[
(\gen{S}^{+}_{1}F_{n-1})\ifarxiv(1,\ldots,n)\fi
:=\int d^{4|4}\Lambda\, S^+(1,2,\bar\Lambda)\,F_{n-1}(\Lambda,3,\ldots,n).
\]
Note that it shifts all the legs $2,\ldots n-1$ of $F_{n-1}$ by one index to $3,\ldots,n$.
We then use the cyclic shift operator 
$(\mathcal{U}_nF_n)(1,\ldots,n):=F_n(2,\ldots,n,1)$
to bring the correction term into all possible places\ifarxiv\else\ \fi%
\footnote{Note that the shift operators
$\mathcal{U}_n$ and $\mathcal{U}_{n-1}$ 
act on two different spaces. 
Thus $\gen{S}^{+}_{k}$ is not periodic: 
$\gen{S}^{+}_{k+n}=\gen{S}^{+}_{k}\mathcal{U}^{-1}_{n-1}$.
In physical situations we act only on cyclic states where $\mathcal{U}_{n-1}\simeq 1$
such that $\gen{S}^{+}$ preserves cyclicity.}
\[
\gen{S}^+
=
\sum_{k=1}^n
\gen{S}^{+}_{k},
\qquad
\gen{S}^{+}_{k}:=
\mathcal{U}^{k-1}_{n}
\gen{S}^{+}_{1}
\mathcal{U}^{1-k}_{n-1}.
\]

For our new symmetry generator $\genY{B}$ 
we propose the following correction term $\genY{B}^+$ 
\begin{align}
\genY{B}^+
=&
\sum_{k=1}^{n-1}\sum_{j=k+1}^n 
\lrbrk{
\ifarxiv\else\begin{array}{l}\fi
\gen{Q}_k^{\alpha b} \gen{S}^+_{j,\alpha b} 
-\bar{\gen{Q}}_{k,b}^{\dot{\alpha}}\bar{\gen{S}}^{+,b}_{j,\dot\alpha}
\ifarxiv\else\eqcr[0.3em]\qquad\fi
-\gen{Q}_j^{\alpha b} \gen{S}^+_{k-1,\alpha b} 
+\bar{\gen{Q}}_{j,b}^{\dot{\alpha}}\bar{\gen{S}}^{+,b}_{k-1,\dot\alpha}
\ifarxiv\else\end{array}\fi\label{shift}
}.
\end{align}
Note the shift of argument for $\gen{S}^+$
as compared to \eqref{eq:Bhdef}
when $\gen{Q}$ acts further to the right.

As a first check we consider cyclicity of $\genY{B}+\genY{B}^+$ 
(supposing we act on cyclic functions)
\begin{align}
\label{eq:cyclexact}
(\mathcal{U}_n-1)(\genY{B}+\genY{B}^+)
=&
-2\gen{Q}_1^{\alpha b}
\bigbrk{\gen{S}_{\alpha b}+\gen{S}^+_{\alpha b} }
\ifarxiv\else\nln\eqtab\fi
+2\bar{\gen{Q}}_{1,b}^{\dot{\alpha}}
\bigbrk{\bar{\gen{S}}_{\dot{\alpha}}^b+\bar{\gen{S}}^{+,b}_{\dot\alpha}}
\\&
+\gen{Q}^{\alpha B} 
\bigbrk{2\gen{S}_{1,\alpha b}+\gen{S}^+_{0,\alpha b} +\gen{S}^+_{1,\alpha b} }
\ifarxiv\else\nln\eqtab\fi
-\bar{\gen{Q}}_{b}^{\dot{\alpha}}
\bigbrk{2\bar{\gen{S}}_{1,\dot{\alpha}}^b+\bar{\gen{S}}^{+,B}_{1,\dot\alpha}+\bar{\gen{S}}^{+,b}_{0,\dot\alpha}}.
\nonumber
\end{align}
The $\gen{Q}$'s anticommute exactly with the $\gen{S}^+_k$'s
\cite{Bargheer:2009qu}, therefore the action of $\genY{B}+\genY{B}^+$ is cyclic.
Interestingly, only the combination of 
$\genY{B}$ and $\genY{B}^+$ is cyclic
because only the combination 
$\gen{S}+\gen{S}^+$
annihilates amplitudes exactly.

More importantly, we can show exact invariance of MHV amplitudes. 
To show \eqref{eq:ExactSym} we note the action 
of $\bar{\gen{S}}^{+}$
\begin{align}
\bar{\gen{S}}^{+,b}_{k,\dot\alpha}\Amp_{n-1,2}
&=
2\pi\varepsilon_{\dot\alpha\dot\beta}
\lrbrk{\bar\lambda_{k}^{\dot\beta}\eta_{k+1}^b
-\bar\lambda_{k+1}^{\dot\beta}\eta_{k}^b}
\ifarxiv\else\nln\eqtab\qquad\cdot\fi
\delta^2(\sprod{k}{k+1})
\frac{\delta^8(Q)\,\delta^4(P)}{\prod_{i\neq k}\sprod{i}{i+1}}\,.
\end{align}
By construction almost all distributional terms cancel.
Only at the boundary there are some residual terms
for which we need some identities to
show full cancellation
\begin{align}
0=&\mathrel{}\varepsilon_{\dot\alpha\dot\beta}
\bar{\gen{Q}}_{1,b}^{\dot{\alpha}}
\bar\lambda_{1}^{\dot\beta}
=
\varepsilon_{\dot\alpha\dot\beta}
\bar{\gen{Q}}_{1,b}^{\dot{\alpha}}
\bar\lambda_{2}^{\dot\beta}\eta_{1}^b
\delta^2(\sprod{1}{2})
\ifarxiv=\else\nln=\eqtab\mathrel{}\fi
\bar{\gen{Q}}_{b}^{\dot{\alpha}}
\delta^8(Q)\delta^4(P)
.
\end{align}

\section{Conclusions and Outlook}

In view of the conjectured integrability for the planar S-matrix
of $\superN=4$ SYM, and its many useful applications, 
it is extremely important to understand the underlying symmetries.
In this letter, we have proposed that there exists an exact symmetry 
besides the established Yangian algebra $\grp{Y}(\alg{psu}(2,2|4))$. 
This Yangian-like symmetry generator $\genY{B}$ 
is the level-one recurrence of the hypercharge $\gen{B}$,
both of which are included in the bigger algebra $\grp{Y}(\alg{u}(2,2|4))$.
Now curiously, the novel $\genY{B}$ appears to be a symmetry
whereas $\gen{B}$ clearly is none. 
This leads to an intriguing structure of the symmetry algebra 
somewhere in between $\grp{Y}(\alg{psu}(2,2|4))$
and $\grp{Y}(\alg{u}(2,2|4))$. 

We have shown explicitly that the bonus Yangian symmetry $\genY{B}$ is a symmetry of all tree-level amplitudes,
and argued that the symmetry is exact in a distributional sense, 
at least for MHV amplitudes.
Cyclicity of color-ordered amplitudes is respected \eqref{eq:cyclfree,eq:cyclexact}. 
All this, in conjunction with the invariance of the Grassmanian integral, 
leads to the conclusion that $\genY{B}$ stands a good chance of being 
a symmetry of loop amplitudes\ifarxiv.\else\ \fi
\footnote{Loop integrals break invariance of the S-matrix, 
yet in a controllable fashion. 
We expect $\protect\genY{B}$ to behave like all previously known superconformal and Yangian generators.
}%
\ifarxiv\else. \fi
Similarly the question arises whether $\genY{B}$ is a symmetry of the (bulk) 
higher-loop spin chain Hamiltonian for planar anomalous dimensions 
of local operators, cf.\ the reviews \cite{Beisert:2004ry,Beisert:2010jr}.

Notably, the new symmetry is stronger than the
dual symmetries.
Together with the ordinary superconformal symmetries
we can generate all previously known symmetries of the S-matrix
including the dual superconformal ones
(this also holds when the correction term $\genY{B}^+$ is considered) via
\[
\comm{\genY{B}}{\gen{Q}}=+\genY{Q},\qquad
\comm{\genY{B}}{\gen{S}}=-\genY{S},\qquad\ldots
\]
Conversely, the ordinary and dual superconformal symmetries
only close onto the Yangian $\grp{Y}(\alg{psu}(2,2|4))$.
As an outer automorphism our symmetry can never be generated in this fashion.
Therefore one might wonder if $\genY{B}$ actually yields
stronger constraints for the S-matrix than the dual symmetries:
Abstractly this is to be expected, but potentially the S-matrix is special 
and invariance under $\genY{B}$ is automatic,
cf.\ \cite{Drummond:2010uq}.
Invariance of the proposed all-loop \emph{integrand} \cite{ArkaniHamed:2010kv}
in fact follows from invariance of the Grassmannian integral
by construction.

It would also be desirable to shed some light on
the (geometric) transformation induced by $\genY{B}$ which 
is at the first level of the Yangian in both the original 
and dual picture of the S-matrix, i.e.\ it is simple in neither picture.

To finish, we comment on scattering amplitudes of $\superN=6$ super Chern-Simons theory, 
which enjoy a similar Yangian symmetry \cite{Bargheer:2010hn}.
Its Yangian $\grp{Y}(\alg{osp}(6|4))$
does not admit an outer automorphism, 
however, the action of the generator $\genY{R}$ is somewhat reminiscent of
our bonus Yangian symmetry $\genY{B}$.

\ifarxiv\paragraph{Acknowledgments.}\else\begin{acknowledgments}\fi
We would like to thank L.\ Ferro, T.\ Matsumoto, T.\ McLoughlin and J.\ Plefka for useful discussions.
The work of N.B.\ is supported in part by the German-Israeli Foundation (GIF).
\ifarxiv\else\end{acknowledgments}\fi

\ifarxiv
\bibliographystyle{nb}
\fi
\bibliography{old,bonussymmetry}

\end{document}